\documentclass[a4paper,11pt]{article}   

\usepackage{authblk} 
\usepackage{xspace}
\usepackage{plain}
\usepackage{pdfpages}
\usepackage{tikz}
\usepackage{caption}
\usepackage{subcaption}
\usetikzlibrary{mindmap}
\usepackage[margin=1.0in]{geometry}

\newcommand{\mts}{\emph{mts}\xspace}
\newcommand{\lrs}{\emph{lrs}\xspace}

\newcommand{\mplrs}{\emph{mplrs}\xspace}

\newcommand{\mait}{\emph{mai20}\xspace}

\usepackage{todonotes}
\usepackage{amsmath}
\usepackage{amssymb}
\usepackage{latexsym}
\usepackage{graphicx}
\usepackage{color}
\usepackage{hyperref}
\usepackage{slashbox}
\usepackage{algorithm}
\usepackage{algpseudocode}

\definecolor{darkblue}{rgb}{0,0,0.6}

\include{./mathlib}
\include{./macros}

\begin{document}

\title{An analysis of budgeted parallel search on conditional Galton-Watson trees}

\title{An analysis of budgeted parallel search on conditional Galton-Watson trees}
\author[1,2]{David Avis}
\author[1]{Luc Devroye}
\affil[1]{School of Computer Science,
          McGill University, Montr{\'e}al, Qu{\'e}bec, Canada}
\affil[2]{School of Informatics, Kyoto University, Kyoto, Japan}

\maketitle

\begin{abstract}
Recently Avis and Jordan have demonstrated the efficiency of a simple technique
called budgeting for the parallelization of a number of tree search algorithms.
The idea is to limit the amount of work that a processor performs before
it terminates its search and returns any unexplored nodes to a master process.
This limit is set by a critical budget parameter which determines the overhead
of the process. In this paper we study the behaviour of the budget parameter
on conditional Galton-Watson trees obtaining asymptotically tight bounds on this overhead.
We present empirical results to show that this bound is surprisingly accurate
in practice.
\end{abstract}

\section{Introduction}
The majority of algorithms have been designed, analyzed and implemented to run
on single core processors.
While multicore hardware is now ubiquitous these algorithms profit little, if any,
from the additional processing power.
On the other hand, the study of parallel algorithms also has a long history.
Issues involved are complex and include architecture design, communication, data sharing,
interrupts, deadlocks, load balancing, and the distinction between shared memory
and distributed computing. For a comprehensive reference, see
Mattson (2004). This complexity presents serious challenges to developing theoretical analyses
to explain successful empirical results. Indeed many recent highly publicized
computational success stories, such as Computer Go, Machine Learning and Deep
Learning for AI, make massive use of parallel computation but have little
or no theoretical foundation for the underlying algorithms. 

In a recent paper, McCreesh and Prosser (2015) attempt to fill this gap in the context
of parallel branch and bound algorithms for the maximum clique problem. They show
that the interaction of the search order and the most likely location of solutions
is often the dominant consideration in load balancing for these types of problems.
This leads to a new work-splitting technique which gives good scalability.
In their conclusions they recall the following prescient remark of Hooker (1995):
\begin{quote}
{\footnotesize
Based on one's insight into an algorithm, for instance, one might expect good performance to depend on
a certain characteristic. How to find out? Design a controlled experiment that checks how the presence
or absence of this characteristic affects performance. Even better, build an explanatory mathematical
model that captures the insight, as is done routinely in other empirical sciences, and deduce from it
precise consequences that can be put to the test.
}
\end{quote}

The purpose of this paper
is to make a modest additional contribution along these lines
by giving a theoretical analysis of the overhead produced by a
simple and
effective 
scaleable parallelization technique,
described by
Avis and Jordan (2015, 2016), 
that is  applicable to a certain class of tree search algorithms. This class includes reverse search
which has been used for a large number of combinatorial and geometric enumeration problems.
Trees in this class must be definable by an oracle which, given any node in the tree, gives
the children of that node (if any) in an arbitrary but fixed order.
Traversals of trees in this class can be performed in parallel by assigning
independent subtrees to different processors. When the underlying tree is unbalanced, the main problem
again becomes one of load balancing.
The tree traversal problem differs significantly from
branch and bound, where the goal is to explore as little of the tree as possible
until a desired node is found. The bounding step removes subtrees from consideration and
this step depends critically on what has already been discovered. Hence the order of
traversal is crucial and the number of nodes evaluated varies dramatically depending on this
order, as reported by McCreesh and Prosser (2015). 
Sharing of information is critical to the success of parallelization in branch and bound.
These issues do not occur when the entire tree must be traversed, hence simpler
parallelization techniques are possible.

Avis and Jordan approach the load balancing problem by using a simple technique called
{\em budgeting}.
A search of a subtree of the original tree
terminates after a given number of nodes, called the {\em budget},
have been generated
and returns all unexplored nodes to a master process which stores them on a job list.
In a multiprocessor setting, the master process assigns nodes from the job list
to available processors, called workers, that proceed in parallel, repopulating the list
whenever their budget is reached. In their 2015 paper they apply this technique to 
parallelize \lrs, a reverse search based code for the vertex/facet enumeration problem for convex polyhedra. 
Experimental results
showed near linear speedups when using up to several hundred processors.
In the second paper the authors describe a generic 
common wrapper that can be used with a wide variety of legacy codes.
They gave applications to other reverse search codes and, more recently,
to satisfiability testers,
again obtaining substantial speedups.
The resulting software, called \mts\footnote{\url{http://cgm.cs.mcgill.ca/~avis/doc/tutorial.html}},
is freely available.
Budgeting may be seen as a very simple form of {\em job stealing}, as described by Blumofe
and Leiserson (1999). However in job stealing individual workers maintain
their own job lists and poll each other as necessary to obtain additional work,
requiring communications between workers, interrupt handling, deadlock avoidance and so on.

Budgeting has several very practical advantages. Firstly it does not require
communication between workers and they do not need to process interrupts.
This means that parallelism does not have to be added to the existing legacy code.
Secondly large subtrees are automatically broken up, since they exceed the budget,
whereas small subtrees are not, since they do not. Thirdly the master can dynamically
control the size of the job list by varying the size of the budget when assigning
work. Finally it is easy to checkpoint the process
since all workers return to the master regularly due to the budget restriction. It is simply
a matter of waiting for all workers to return and then outputting the remaining job list.
On the other hand budgeting introduces two competing forms of overhead. One is the cost of
restarting jobs on the master's list, which is directly proportional to the total size
of the list. The other is the cost of idle workers if the list becomes empty.
It is therefore very important to determine
how the budget parameter affects the size of the job list.

The goal of this paper is to analyze the budgeting method when
it is applied to critical
Galton-Watson trees. 
These trees provide a challenge to efficient parallel search since they are very unbalanced:
it is well-known, see, e.g., Flajolet and Odlyzko (1982), that the height of such trees is of the order
or the square root of their size.
For this class of trees we study the critical budget parameter and how it determines the
size and evolution of the job list. 
In Section 2 we formally introduce budgeting and give
an explicit example of how it can be implemented in a depth first search setting.
Next, in Section 3 we introduce Galton-Watson trees and observe that they include quite a large
number of other classes of random trees.
We then state our main result, a tight estimate on the overhead
produced by budgeting for traversing trees in our family. 
In Section 4 we describe a random
walk which is closely related to the evolution of the list of unexplored nodes
produced by budgeting.
This is followed in Section 5 by a proof of our main result.
Finally in Section 6
we make use of the \mts framework to empirically test our estimates on
a sample of large Galton-Watson trees. We also give a discussion of how our main
result can be used in a multiprocessor setting in order to choose an efficient budget parameter.

\section{Budgeted tree search}
\label{budget}
We are given a tree $T$ by its root, an upper bound $\Delta$ on the number of children of any node,
and an adjacency oracle $Adj(v,j)$. The adjacency oracle takes a tree node $v$
and an index $j \in \{0,1,2,...,\Delta-1\}$.
For each value of $j$ in its range, $Adj(v,j)$ either returns $null$ or gives a child of $v$. Each
child is given exactly once. Our goal is to visit all of the nodes in $T$.
For concreteness in this section we will describe depth first search (DFS) although
the budgeting technique can be applied to breadth first search or other search strategies.

A budgeted DFS is initiated from tree node
$start\_vertex$, initially set to be the root, and  proceeds in a depth first fashion
outputting each node as it is generated. We use two stacks, $stack\_v$ to
store vertices and $stack\_j$ to store indices.
We also specify an integer budget parameter $b \ge 1$ which causes the
tree search to be terminated when $b$ nodes have been generated. At this point
the most recently generated vertex $v$ and all of the unexplored siblings along the return path 
from $v$ to $start\_vertex$ are
returned with the flag {\em unexplored=true}. All previously generated nodes were output with
the flag {\em unexplored=false}.
The pseudo-code is shown as Algorithm \ref{alg:bdfs}.
Note that that $start\_vertex$ is not output so this should be done
in the calling program. We exploit this property later.

\begin{algorithm}[htb]
\begin{algorithmic}[1]
\Procedure{bdfs}{$start\_vertex$, $\Delta$, $Adj$, $b$ }
        \State $j \gets 0~~~v \gets start\_vertex~~~count \gets 0 ~~~depth \gets 0$
        \Repeat
        \State $unexplored \gets false$
        \While {$j < \Delta$ {\bf and} $unexplored = false$ }
                \State $j \gets j+1$
                \State $push(stack\_v,v)$
                \State $push(stack\_j,j)$
                        \State $v \gets Adj(v,j)~~~~~$
                        \State $depth \gets depth + 1~~~~count \gets count+1$
                        \If {$count \ge b$}  \Comment{budget is exhausted}
                            \State $unexplored \gets true$
                        \EndIf
                        \State {\bf output} $(v,unexplored)$
                        \If {$count < b$}  \Comment{continue down tree}
                            \State $j \gets 0$
                        \EndIf
        \EndWhile
        \If {$depth > 0$}   \Comment{backtrack step}
                \State $v \gets pop(stack\_v)$
                \State $j \gets pop(stack\_j)$
                \State $depth \gets depth - 1$
        \EndIf
        \Until {$depth = 0$ {\bf and} $j=\Delta$}
\EndProcedure
\end{algorithmic}
\caption{Budgeted depth first search}
\label{alg:bdfs}
\end{algorithm}

Consider the tree in Figure \ref{fig:example} which has 25 nodes, $\Delta=5$ and is
rooted at vertex 0.
For convenience the nodes are numbered 0,1,...,24 in depth first search order
but this is in no way essential. 
If we set the parameter $b$ to be 25 or greater, all nodes are output 
in this order with {\em unexplored=false}.
Suppose we set $b=13$. Firstly nodes 1,...,12 are output in order
with {\em unexplored=false}.
Then nodes 13,15,16,18,22 are output with {\em unexplored=true}.
On the other hand, if we set $b=8$ then nodes 1,2,...,7 are output with {\em unexplored=false}
and nodes 8,9,10,11,15,16,18,22 are output with {\em unexplored=true}.

\begin{figure}[htbp]
\centering
\resizebox{0.5\textwidth}{0.3\textwidth}{%
\begin{tikzpicture}[grow cyclic, align=flush center,
    level 1/.style={level distance=3cm,sibling angle=90},
    level 2/.style={level distance=1.6cm,sibling angle=45}]
\node{$0$}
 child { node {$1$}
         child { node {$2$} }
         child { node {$3$} }
         child { node {$4$} }
         child { node {$5$} }
         child { node {$6$} }
       }
 child { node {$7$}
         child { node {$8$} }
         child { node {$9$} }
         child { node {$10$} }
         child { node {$11$}
             child { node {$12$} }
	     child { node {$13$} 
	           child { node {$14$} }
                   }
              }
         child { node {$15$} }
         child { node {$16$}
              child { node {$17$} }
              }
       }
 child { node {$18$}
         child { node {$19$} }
         child { node {$20$ } }
         child { node {$21$} }
       }
 child { node {$22$}
         child { node {$23$} }
         child { node {$24$} }
       };
\end{tikzpicture}
}
\caption{Tree rooted at vertex 0 with 25 nodes and $\Delta=6$}
\label{fig:example}
\end{figure}
We will store the unexplored nodes in a list $L$ which, for DFS,
would be implemented as a stack. To complete
the DFS of the tree we search the subtrees rooted by nodes in $L$ using 
Algorithm \ref{alg:bdfs} repeatedly. Note that each
vertex in $L$ now becomes a $start\_vertex$ and if a budget constraint $b$ is applied
additional nodes may be added to $L$.
If $L$ is empty when Algorithm \ref{alg:bdfs} terminates the tree has been completely output.
Since the root nodes are not output we will obtain a duplicate free list of all the nodes
of the tree by concatenating outputs.
Note that if the order of output is not important, the subtrees rooted by nodes in $L$ can be explored
in any order and in parallel. Furthermore, if $b=1$ then all nodes in $T$ will be returned to $L$.
Alternatively, if $b > |T|$ then $L$ remains empty as the tree is explored with a single
call to Algorithm \ref{alg:bdfs}. 

A successful parallelization of Algorithm \ref{alg:bdfs} depends on controlling the job list $L$.
On one hand, if the budget parameter $b$ is too small, then $L$ tends to become large
causing considerable overhead in restart costs. If $b$ is too large, then $L$ becomes smaller,
reducing overhead,
but may become empty before the entire tree has been searched. This causes processors to become
idle and reduces the speedup that can be obtained. 
We are interested in how $L$ evolves over time.

\begin{figure}[h!tb]
\centering
\begin{subfigure}[b]{0.47\textwidth}
 \centering
 \resizebox{\textwidth}{!}{
 \includegraphics[width=\textwidth]{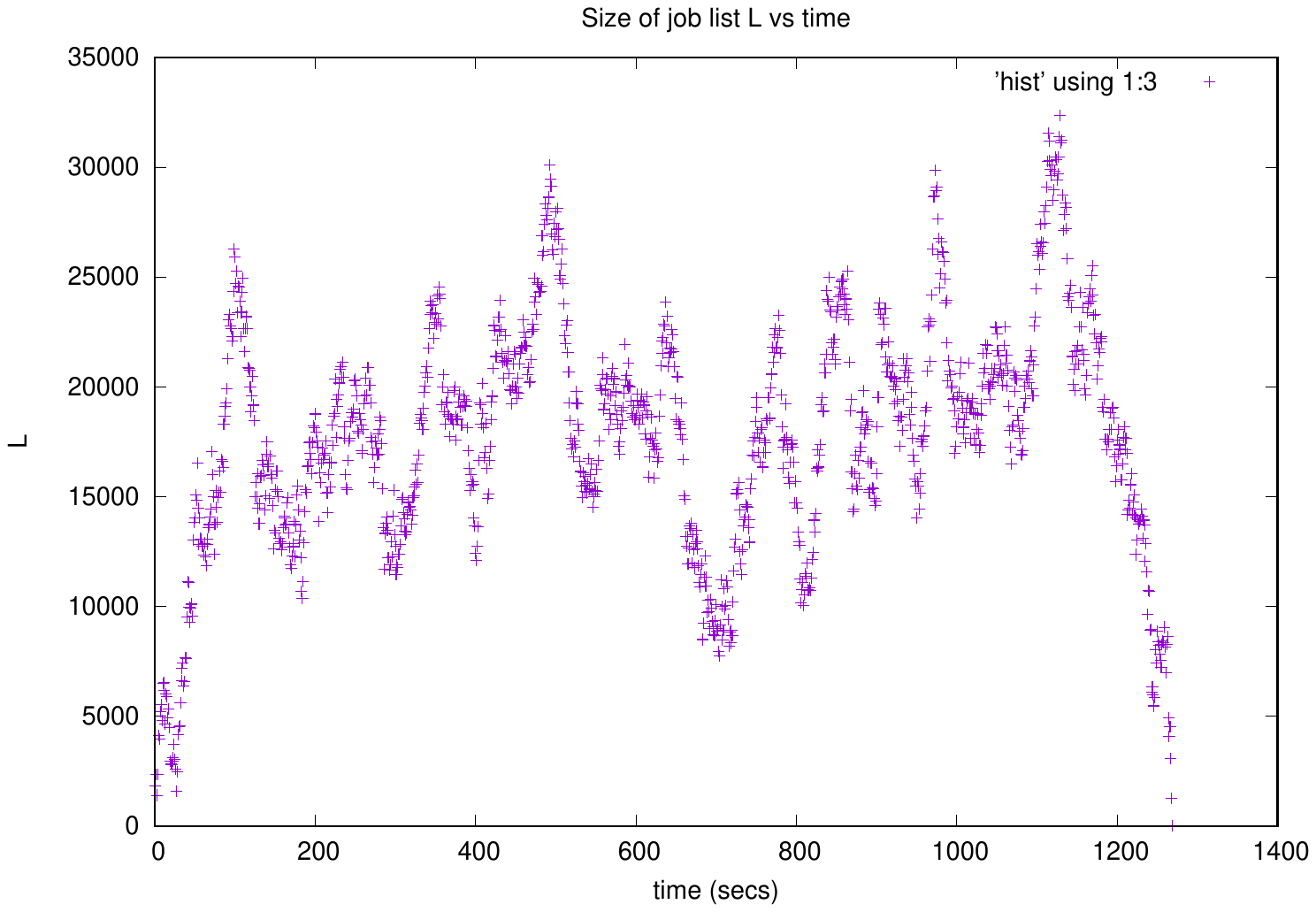}
 }
 \caption{$b=50,~R_n=52,454,166$}
 \label{subfig:50}
\end{subfigure}
\begin{subfigure}[b]{0.47\textwidth}
 \centering
 \resizebox{\textwidth}{!}{
 \includegraphics[width=\textwidth]{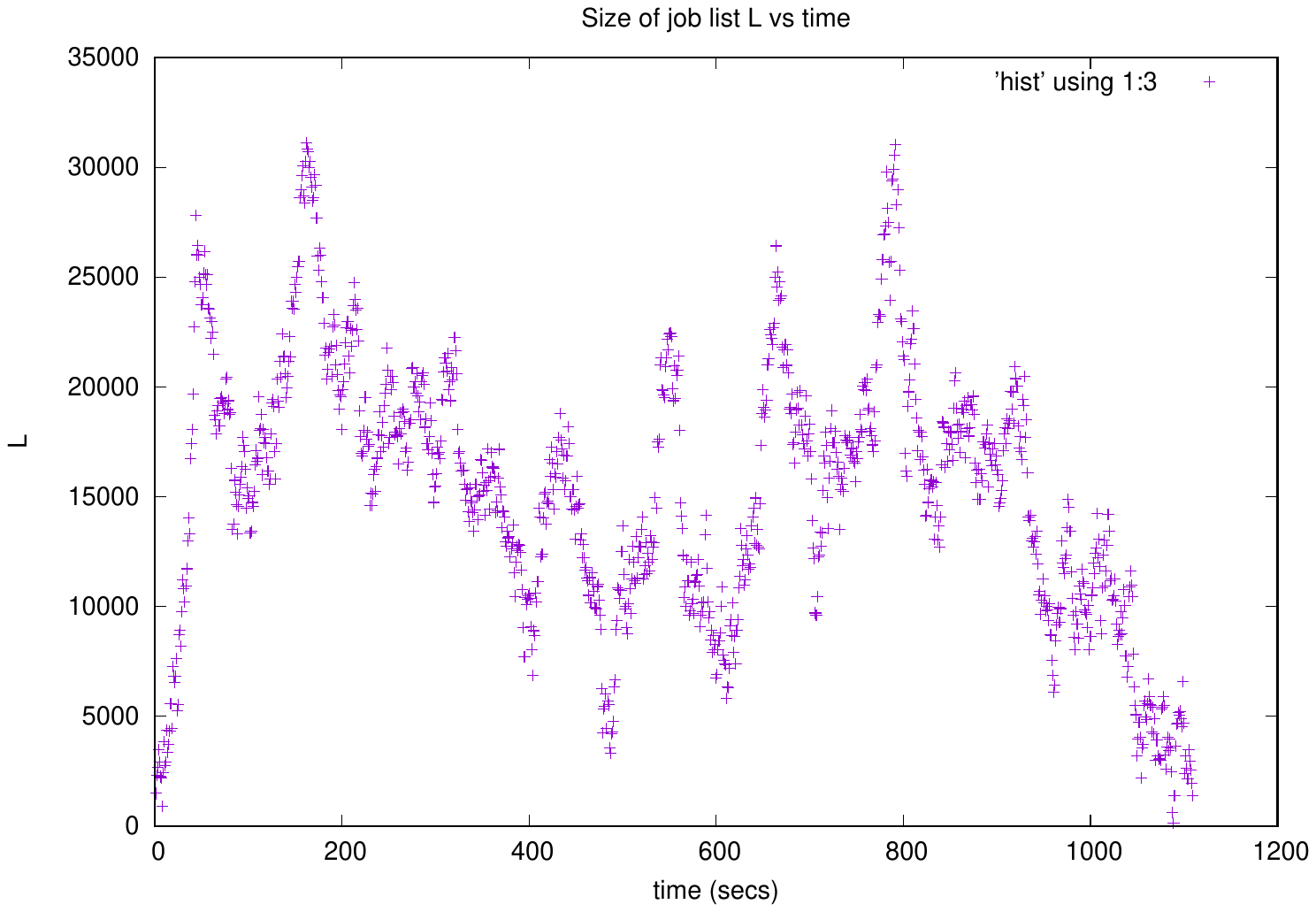}
 }
 \caption{$b=500,~R_n=17,384,235$}
 \label{subfig:500}
\end{subfigure}
\begin{subfigure}[b]{0.47\textwidth}
 \centering
 \resizebox{\textwidth}{!}{
 \includegraphics[width=\textwidth]{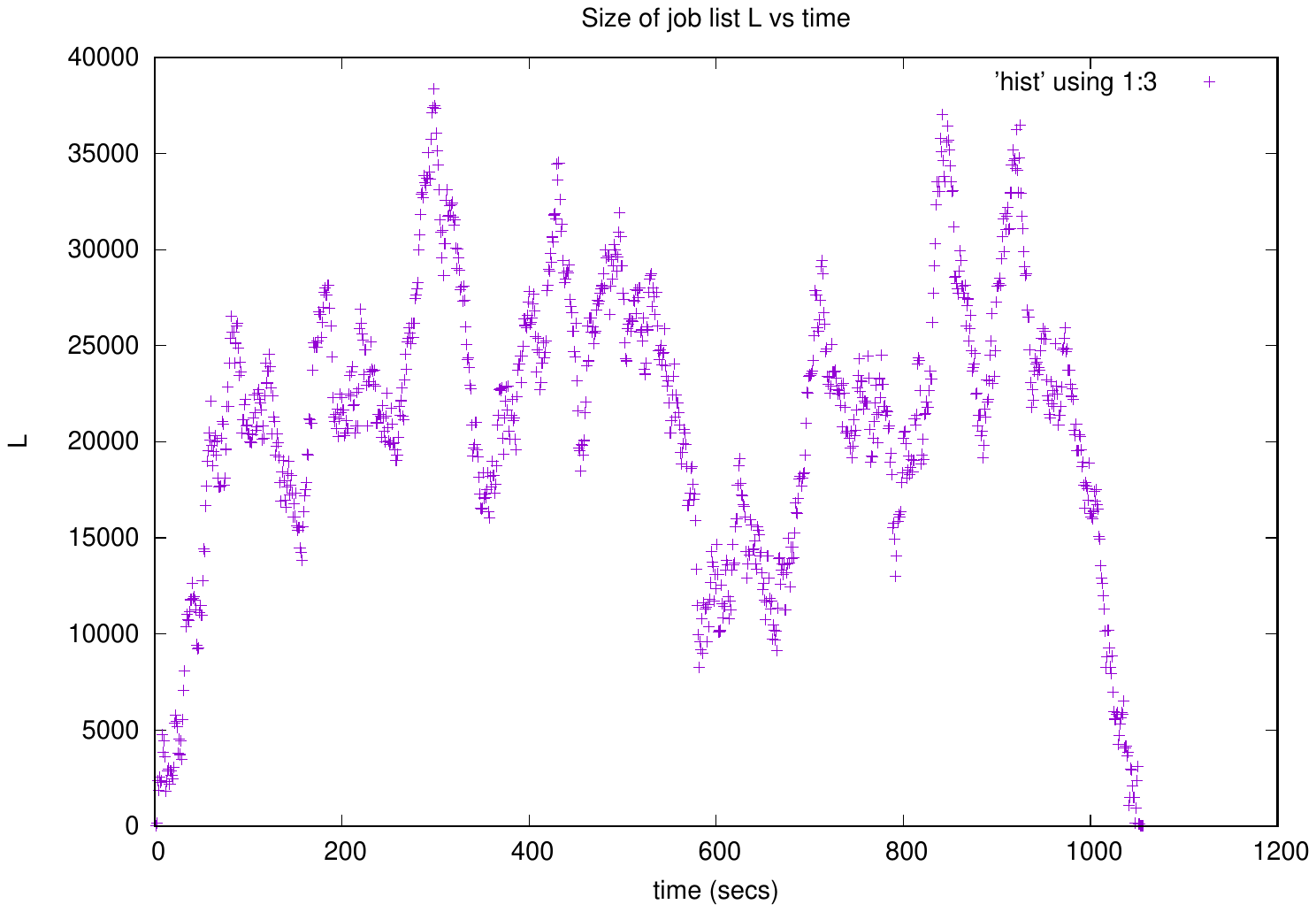}
 }
 \caption{$b=5000,~R_n=5,584,327$}
 \label{subfig:5000}
\end{subfigure}
\begin{subfigure}[b]{0.47\textwidth}
 \centering
 \resizebox{\textwidth}{!}{
 \includegraphics[width=\textwidth]{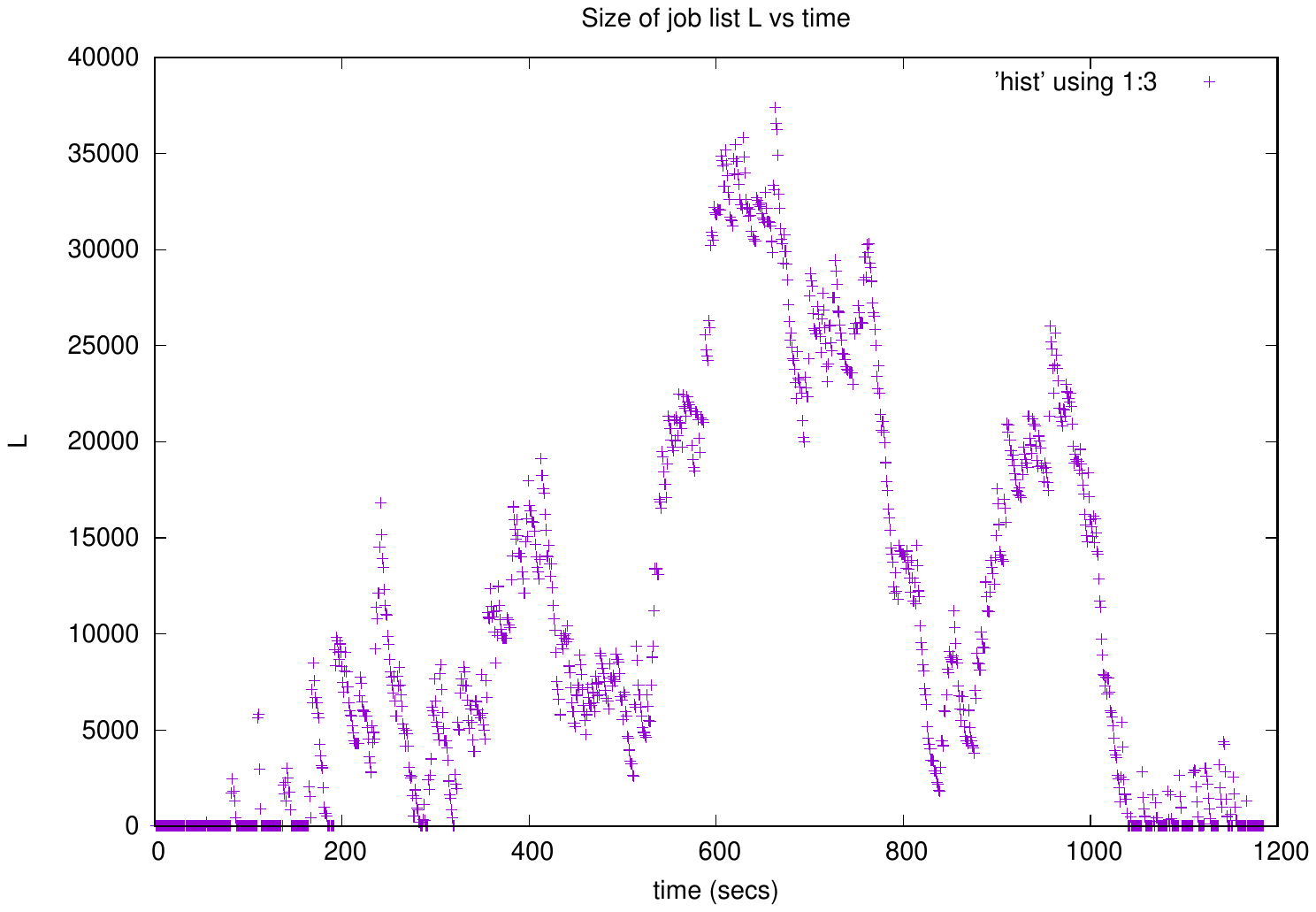}
 }
 \caption{$b=500,000,~R_n=572,507$}
\label{subfig:500000}
\end{subfigure}
\caption{Evolution of the job list for $T_{10}$: budget=$b$, total restarts=$R_n$, \mait, 16 cores}
\label{fig:hist_sample}
\end{figure}

Let $L_b$ be the job list generated by repeatedly applying Algorithm \ref{alg:bdfs} to a tree $T$ 
for a given budget $b$ and let $R_n$ denote the total number of nodes added 
to the list during a complete
search of $T$. 
We can observe the size of $L_b$ each time
Algorithm \ref{alg:bdfs} is called.
For example, again referring to Figure \ref{fig:example}, Algorithm \ref{alg:bdfs}
is called a total of 6 times if we keep $b=13$ throughout. 
$|L_{13}|=5$ and the size of $L_{13}$ is respectively
0,4,3,2,1,0 for these 6 times. 

For a larger example consider Figure \ref{fig:hist_sample}. This is based on visiting all the nodes
of the tree $T_{10}$ which has $\Delta=10$ and 264,933,315
nodes.\footnote{All computational results in the paper were obtained on mai20 at Kyoto University: 2x Xeon E5-2690 (10-core 3.0GHz), 20 cores, 128GB memory, 3TB hard drive}
Its construction is described in
detail in Section 6. The plots were obtained
by running Algorithm \ref{alg:bdfs} in parallel on 16 cores using \mts version 0.1 and different budget
sizes $b$. 

The first thing to observe is that, as expected, the number $R_n$ of jobs returned to $L$
is inversely proportional to $b$. 
By comparing
the results for
$b=50, 5000, 500000$ we see that as the budget scales up by a factor of 100, the size of
$R_n$ scales down by roughly a factor of 10, the square root of the scale factor. 
In the following sections we show that this is
typical behaviour for Galton-Watson trees, of which $T_{10}$ is a sample.

For $b=50$ and $b=500$ we have the desired behaviour for $L$:  
it quickly reaches a reasonable size (compared to the total tree size) and remains non-empty until the end 
of the run, implying all processors are busy. For $b=5,000$ the job list comes dangerously close to being
empty before the run ends and for the extreme case $b=500,000$ it is empty for several time intervals.
All things being equal a small budget would seem optimum. However in practical applications, 
such as vertex enumeration, the
startup cost for a job taken from $L$ can be very significant. 
A budget small enough to keep a non-empty job list yet large enough to prevent too many restarts
leads to optimum performance. 

Before proceeding we note two important points. Firstly the total size of the job
list is independent of the number of workers assigned. Indeed a single worker
repeatedly executing Algorithm \ref{alg:bdfs} produces the same number of jobs as a thousand
workers in parallel. Secondly, if the job list stays well populated for the duration of the run
then workers are never idle. If the jobs on the list are independent, as they
are in a Galton-Watson tree, then each worker behaves 
in the same way in probability.
It is therefore sufficient to study the behaviour of a single worker applying
Algorithm \ref{alg:bdfs} repeatedly until the job list is empty.

\section{The probabilistic model and our main result}
\label{pm}
\begin{figure}
 \centering
 \includegraphics[width=\textwidth]{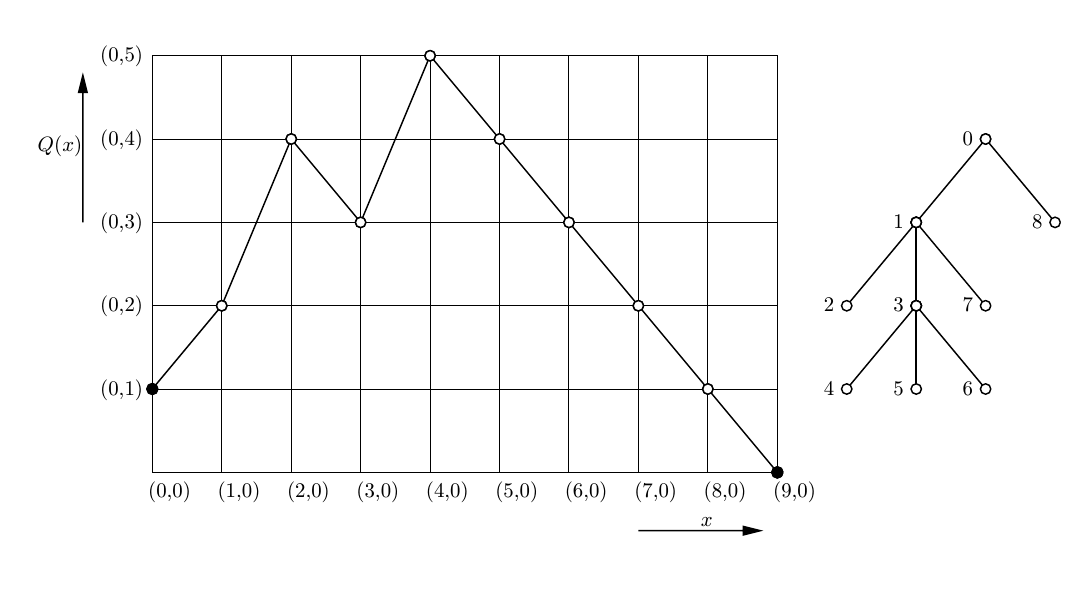}
 \caption{The Q-process derived from a tree}
 \label{Q-process}
\end{figure}

\begin{plain}
\input{luc_defs.tex}

A Galton-Watson (or Galton-Watson-Bienaym\'e) tree (see Athreya and Ney, 1972)
is a rooted random ordered tree. Each node independently 
generates a random number of children drawn from a fixed offspring distribution $\xi$.
The distribution of $\xi$ defines the distribution of $T$,
a random Galton-Watson tree.
In what follows, we are mainly interested in critical Galton-Watson trees, i.e., those
having $\EXP \{ \xi = 1 \}$, and $\PROB \{ \xi = 1 \} < 1$. 
In addition, we assume that the variance of $\xi$ is finite (and hence,
nonzero).

Moon (1970) and Meir and Moon (1978) defined
the {\it simply generated trees} as ordered labelled trees
of size $n$ that are all equally likely given 
a certain pattern of labeling for each node of a given degree.
The most important examples include the Catalan trees (equiprobable
binary trees), the equiprobable $k$-ary trees,
equiprobable unary-binary trees (ordered trees with up to two children),
random Motzkin  trees,
random planted plane trees (equiprobable
ordered trees of unlimited degrees) and Cayley trees
(equiprobable unordered rooted trees).
It turns out that all these trees can be represented
as critical Galton-Watson trees conditional on their size, $n$,
a fact first pointed out by Kennedy (1975),
and further developed by Kolchin (1980, 1986) and others.
So, let $T_n$ be a Galton-Watson tree conditional on its size being $n$.
For example, when
$\xi$ is $0$ or $2$ with probability $1/4$, and $1$ with probability $1/2$,
we obtain the uniform binary (Catalan) tree.
Uniformly random full binary trees are obtained by setting
$\PROB \{ \xi = 0 \} = \PROB \{ \xi = 2 \} = 1/2$.
A uniformly random $k$-ary tree has its offspring distributed
as a binomial $(k, 1/k)$ random variable.
A uniform planted plane tree is obtained for the geometric law 
$\PROB \{ \xi = i \} = 1/2^{i+1}$,
$i \ge 0$.
When $\xi$ is Poisson of parameter $1$, one obtains 
(the shape of) a random rooted labeled (or Cayley) tree.
For $\xi$ uniform on $\{ 0, 1 , 2, \ldots, k  \}$, $T_n$ is like
a uniform ordered tree with maximal degree of $k$.
All such trees can be dealt with at once in the
Galton-Watson framework.

Recall that $R_n$ is the number of restarts generated
when Algorithm \ref{alg:bdfs} is used repeatedly to explore a tree $T$ with $n$
nodes using a budget $b$. Our main result is:

\proclaim Theorem 1.
If $T_n$ is a Galton-Watson tree of size $n$ determined by
$\xi$, where $\EXP \{ \xi \} = 1$ and $0 < \Var \{ \xi \} \isdef \sigma^2 < \infty$,
then
$$
{R_n \over n} \to {1 \over \mu_b}
$$
in probability as $n \to \infty$, where
$$
\mu_b \isdef \EXP \{ \min (|T|,b) \},
$$
and $T$ is an unconditional Galton-Watson tree for the same $\xi$.
In addition, as $b \to \infty$,
$$
\mu_b \sim \sqrt{8b \over \pi \sigma^2}.
$$

\bigskip
\section
4~Preliminary results: the random walk view

Let $\xi$ be a random variable representing the number
of children of the root in a critical Galton-Watson
tree: $\EXP \{ \xi \} = 1$, $\PROB \{ \xi = 0 \} > 0$.
Set $X=\xi -1$, and let $X_1, X_2, \ldots$ be a sequence
of i.i.d.\ random variables distributed as $X$.
Define the partial sums $Q(0)=1$, $Q(t) = 1+ \sum_{i=1}^t X_i$.
There is a well-known depth first or preorder construction
of a random Galton-Watson tree which lends itself
well to the study of all properties of randomly selected
nodes (see, e.g., Le Gall, 1989, or Aldous, 1991). 
Nodes in an ordered tree can be encoded
with a vector of child numbers. The root corresponds
to the empty vector. Its children have encodings
$1, 2, \ldots, \xi$. More generally, the $i-th$ child of
a node with label $j$ is encoded $(j,i)$.
A preorder listing of the nodes is nothing but
a lexicographic listing of the node vectors.

We can traverse a random Galton-Watson tree
by visiting nodes in preorder, starting at the
root. To do so, a list $V$ of nodes to
be visited is kept, which is initially of size one
($V$ just contains the root). 
When node $u$ is visited, we consider $\xi_u$,
the number of children of $u$, and remove $u$ from $V$.
Thus, $V$ increases by $\xi_u -1$.
The next node in lexicographic order is taken from $V$,
and the process continues until $V$ is empty.

We denote by $N$ the size of a random Galton-Watson tree.
The size of $V$ after $t$ nodes have been processed is precisely
the $Q$ process introduced above,
$Q(t)$. Thus, $Q(0) = 1$, and
$$
Q(t) = 1 + X_1 + \cdots + X_t ,
$$
where $X_t = \xi_t -1$, and indexing of the nodes is by
their lexicographic rank. An example of the $Q$ process derived from
a tree is shown in Figure \ref{Q-process}. The path in the grid
from vertex (0,1) to
vertex (9,0) is derived from a preorder traversal of the tree to the right.
Except for vertex (9,0), each vertex of the path is of the form $(x,Q(x))$
where $x$ is the tree node to be explored next and $Q(x)$
is the number of unexplored tree nodes we have discovered so far. We start at
(0,1) as $V$ contains only the root 0 and so $V$ has size $Q(0)=1$. 
Since the root has two children the list of unexplored nodes becomes $\{1,8\}$
which has length 2 so we move to vertex (1,2) in the grid.
Tree node 1 has 3 children so the list becomes $\{2,3,7,8\}$ with length 4,
we move to (2,4) and so on. Finally we explore node 8 which has no children,
the unexplored list is empty and we terminate the Q process.

We have the identity
$$
[ N = n ] = [ Q(1) > 0 , Q(2) > 0, \ldots, Q(n-1) > 0, Q(n) = 0 ].
$$
The standard circular symmetry argument for random walks (see, e.g., Dwass, 1968)
shows that
$$
\eqalignno{
\PROB \left\{  N = n \right\}
&= \PROB \left\{ Q(0) = 1 , Q(1) > 0 , Q(2) > 0, \ldots, Q(n-1) > 0, Q(n) = 0 \right\} \cr
&= {1 \over n} \PROB \left\{ Q(n) = 0 \right \} \cr
&= {1 \over n} \PROB \left\{ X_1 + \cdots + X_n = -1 \right \}. \cr}
$$
\end{plain}
\begin{plain}
\input{luc_defs.tex}
The asymptotics for this probability distribution are well-known,
and will be given below.
Let 
$0 < \sigma^2 
\isdef \Var \{ \xi \} > 0$ (which implies $\PROB \{ \xi = 1 \} < 1$),
and let the span $d$ be the greatest common divisor of all $i>0$ for
which $\PROB \{ \xi = i \} > 0$. A Galton-Watson tree
with span $d$ can only have sizes that are $\1modd$.
From Petrov (1975, p.~197) or Kolchin (1986, p.~16, p.~105), we recall that
$$
\lim_{n \to \infty} \sup_{k = \1modd}
\left| \sigma \sqrt{n} \, \PROB \left\{ X_1 + \cdots + X_n = k \right \}
 - {d \over \sqrt{2\pi} } e^{-{k^2 \over 2 \sigma^2 n}} \right|
 = 0, 
$$
while, clearly, $\PROB \left\{ X_1 + \cdots + X_n = k \right \} = 0$ 
if $k \not= \1modd$.
Thus,  along $n \in {\cal N}$,
$$
\PROB \left\{  N = n \right\} \sim { d \over \sigma \sqrt{2 \pi} n^{3/2} },
$$
and, as $n  \to \infty$,
$$
\PROB \left\{  N \ge n \right\} \sim \sqrt{ 2 \over \pi n \sigma^2}.
$$

\proclaim Lemma 1.
Under the conditions of Theorem 1, as $b \to \infty$,
$$
\EXP \{ \min (N,b) \} \sim  
\sqrt{ 8b \over \pi \sigma^2}.
$$

\proof
We have
$$
\eqalignno{
\EXP \{ \min (N,b) \} 
&= \sum_{t=1}^\infty t \PROB \{ \min (N,b) = t \} \cr
&= \sum_{t=1}^b t \PROB \{ N = t \} + b \PROB \{ N >b  \} \cr
&= \sum_{t=1}^b \PROB \{ X_1+\cdots+X_t = -1 \} + b \PROB \{ N >b  \} \cr
&= \sum_{t\le b: t \in {\cal N}} {d+o(1) \over \sigma \sqrt{2 \pi t }} + (1+o(1)) \sqrt{ 2b \over \pi \sigma^2} \cr
&\sim \int_1^b {1 \over \sigma \sqrt{2 \pi t}} \,dt + \sqrt{ 2b \over \pi \sigma^2} \cr
&\sim \sqrt{ 8b \over \pi \sigma^2}.  ~\square \cr
}
$$
\medskip

Consider the $Q$-process conditional on visiting $(x,y)$,
with $0 < x < n$ and $y > 0$. Define
$$
S(x,y) = \min \{ t > x: Q(t) = Q(x)-1 ~|~Q(x)=y \} - x,
$$
where we recall that $Q(t) = 1 + X_1 + \cdots + X_t$.
In other words, $S(x,y)$ is the size of the subtree of
node $x$ in depth first search order given that $Q(x)=y$.
Referring back to the example in Figure \ref{Q-process}
we see that 
$S(0,1)=9,~S(1,2)=7,~S(2,4)=1,~S(3,3)=4,~S(4,5)=1,~S(5,4)=1,~S(6,3)=1,~S(7,2)=1,~S(8,1)=1.$

Note that $S(x,y) = \1modd$.
In addition, since $Q(0)=1$, we must have $t+Q(t)=\1modd$,
so that the only values $(x,y)$ of interest to us have $x+y=\1modd$.
We consider two positive constants $a$, $c$, with $a < 1/2, c < 1$,
and let ${\cal S}(a,c)$ denote the collection of all
random variables $S(x,y)$  with $an \le x \le (1-a)n$,
$c \sqrt{n} \le y \le (1/c) \sqrt{n}$. 
Lemma 2 below shows that within ${\cal S}(a,c)$,
all random variables are close to $N$, the size of
an unconditional Galton-Watson tree. This is an explicit
version of a well-known theorem due to Aldous (1991) (see also
Janson (2012)), which states that if $T^*_n$ is the subtree of a uniform
random node of $T_n$, then $T^*_n$ tends in distribution to an
unconditional Galton-Watson tree $T$ as $n \to \infty$.

\proclaim Lemma 2.
For any fixed $0 < a < 1/2$, $0 < c < 1$ and integer $b > 0$,
$$
\lim_{n \to \infty}
\sup_{1 \le i \le b; i=\1modd}  \,
\sup_{(x,y): \cases{ S(x,y) \in {\cal S}(a,c) &\cr x+y=\1modd &\cr}}
\left| { \PROB \{ S(x,y) = i \} \over \PROB \{ N = i \} } - 1 \right| = 0.
$$
Furthermore,
$$
\lim_{n \to \infty}
\sup_{(x,y): \cases{ S(x,y) \in {\cal S}(a,c) &\cr x+y=\1modd &\cr}} 
\left| \EXP \{ \min(b, S(x,y)) \} - \EXP \{ \min (b, N) \} \right| = 0.
$$

\proof
The last part follows from the first one, so we only consider
$\PROB \{ S(x,y) = i \}$ for $1 \le i \le b$, $i=\1modd$.
We have, for $x+y = \1modd$, $n=\1modd$,
$$
\eqalignno{
\PROB &\{ Q(x)=y, Q(t) > 0 ~\hbox{\rm for all}~x < t < n,   Q(n)=0 \} \cr
&= \PROB \{ Q(x) = y \} \times {1 \over n-x} \PROB \{ X_1 + \cdots + X_{n-x} = -y \} \cr
&= \PROB \{ Q(x) = y \} \times {d+o(1)  \over (n-x) \sqrt{2 \pi (n-x) \sigma^2}} \exp \left( - {y^2 \over 2(n-x)\sigma^2} \right). \cr}
$$
Also,
$$
\eqalignno{
\PROB & \{ Q(x)=y, S(x,y)=i , Q(t) > 0 ~\hbox{\rm for all}~x+i < t < n,   Q(n)=0 \} \cr
&= \PROB \{ Q(x)=y \} \times \PROB \{  N=i \} \times {1 \over n-x-i} \PROB \{ X_1 + \cdots + X_{n-x-i} = -(y-1) \}  \cr
&= \PROB \{ Q(x)=y \} \times \PROB \{  N=i \} \times {d+o(1)  \over (n-x-i) \sqrt{2 \pi (n-x-i) \sigma^2}} \exp \left( - {(y-1)^2 \over 2(n-x-i)\sigma^2} \right). \cr}
$$
Thus, as
$$
\PROB \{ S(x,y) = i \} 
= { \PROB \{ Q(x)=y, S(x,y)=i , Q(t) > 0 ~\hbox{\rm for all}~x+i < t < n,   Q(n)=0 \}
\over
\PROB \{ Q(x)=y, Q(t) > 0 ~\hbox{\rm for all}~x < t < n,   Q(n)=0 \} },
$$
we have
$$
\eqalignno{
{\PROB \{ S(x,y) = i \} \over \PROB \{  N=i \} }
&= (1+o(1)) \left( { n-x \over n-x-i }\right)^{3/2} \exp \left( {y^2 \over 2(n-x)\sigma^2} - { (y-1)^2 \over 2(n-x-i)\sigma^2}  \right) \cr
&= (1+o(1)) \exp \left( {y^2 \over 2(n-x)\sigma^2} - { (y-1)^2 \over 2(n-x-i)\sigma^2}  \right) \cr
&= (1+o(1)) \exp \left( {y^2 - (y-1)^2 \over 2(n-x)\sigma^2} \right) \cr
&= (1+o(1)) \exp \left( {2y-1 \over 2(n-x)\sigma^2} \right) \cr
&= 1+o(1), \cr}
$$
where all the $o(1)$ terms
are uniform over all $i,x,y$ within the given ranges. $\square$
\medskip

The $Q$ process can be viewed as a continuous curve on $[0,n]$
by using linear interpolation between $(t,Q(t))$ and $(t+1,Q(t+1))$.
When properly rescaled, it converges in distribution to 
Brownian excursion.  
 Brownian excursion is 
 Brownian motion $B(t)$ on $[0,1]$
 conditioned to be positive and to take the value 0 at time 1. 
 Alternatively, it is a Brownian bridge process conditioned to be positive. 
Another representation of a Brownian excursion 
$e(t)$ due to Paul L\'evy (and noted by Ito and McKean, 1974) 
is in terms of the last time $\tau_{-}$ that Brownian motion
$B(t)$ hits zero before time 1 and the first time $\tau_{+}$
that Brownian motion $B(t)$ hits zero after time 1:
$$
\left\{ e(t): 0 \le t \le 1 \right\}
\inlaw
\left\{ { |B((1-t)\tau_- + t\tau_+) | \over\sqrt{\tau_+-\tau_-} }: 0 \le t \le 1 \right\}.
$$
Since $B$, and thus $e$, are continuous processes,
we note the following:
\item{(i)}
$\max_{0 \le t \le 1} e(t) $ is a random variable.  In fact,
it has the theta distribution, as proved by 
R\'enyi and Szekeres (1967),
Kennedy (1976),
Flajolet and Odlyzko (1982) and others.
\item{(ii)}
For every $\epsilon \in (0,1/2)$, $\min_{\epsilon \le t \le 1-\epsilon} e(t)$ is a random variable
without an atom at zero (for otherwise, it would
contradict L\'evy's representation).
\item{}

The convergence of partial sums of i.i.d.\ zero mean random variables
with a finite variance to Brownian motion is well-known. As
 a consequence, partial sums conditioned on being positive and
 attaining 0 at the start and end, when suitably rescaled,
 converge to Brownian excursion.
In particular, 
$$
\left( {Q(tn) \over \sigma \sqrt{n}} , 0 \le t \le 1 \right)
\tendsinlaw
( e(t), 0 \le t \le 1 )
$$
in the sup norm metric. In other words, there exists
a sequence of Brownian excursions $e_1, e_2, \ldots$,
such that
$$
\sup_{0 \le t \le 1} \left| {Q(tn) \over \sigma \sqrt{n}} - e_n (t) \right|
\to 0
$$
in probability as $n \to \infty$. 
The work on this convergence of the $Q$ process goes back to
Le Gall (1989, 2005),
Aldous (1991, 1993),
Bennies and Kersting (2000), 
Marckert and Mokkadem (2003),
Duquesne (2003), and others.
A consequence of this and remarks 
(i) and (ii) above is
that for given $\epsilon > 0$ and $\delta > 0$,
we can find $\theta > 0$ such that with probability at least $1-\delta$,
$$
\theta \sqrt{n} \le \min_{\epsilon n \le t \le (1-\epsilon) n} Q(t)
\le \max_{\epsilon n \le t \le (1-\epsilon) n} Q(t) \le {1 \over \theta} \sqrt{n}.
$$
Call this event $G(\epsilon, \theta)$.
\end{plain}

\begin{plain}
\input{luc_defs.tex}

\section
5~Proof of Theorem 1

We are now ready to prove Theorem 1.
Let us denote the sizes of the trees explored by the depth
first search steps by $N_1, N_2, N_3, \ldots$,
where we note that $1 \le N_i \le b$ for all $i$.
The indices of the nodes at which the depth first search
operations are started are denoted by $T_1 < T_2 < \cdots$,
with $T_1 = 0$, $T_{i+1}=T_i+N_i$ for all $i$.
We have
$$
R_n = R = \min \{r > 0: T_r = n-1 \}.
$$
This forces $Q(T_R)=1$,
and thus $Q(T_{R}+1)=0$.  In total, the driving stack has
held $R_n$ nodes.

By duality,
$$
[ R \ge r ] \equiv
[ N_1 + \cdots + N_r < n ].
$$
Recalling the event $G(\epsilon,\theta)$ from the previous section,
and denoting by $I(\epsilon)$
the interval of indices $[\epsilon n, r -\epsilon n]$,
we have
$$
\PROB \{ R \ge r \}
\le \PROB \left\{ \sum_{i \in I(\epsilon)} N_i < n \right\}
= \PROB \left\{ \sum_{i \in I(\epsilon)} (N_i-\EXP N_i) < n - \sum_{i \in I(\epsilon)}  \EXP N_i \right\}.
$$
Note that for $i \in I(\epsilon)$, we have $T_i \ge i \ge \epsilon n$,
and $T_i \le r-\epsilon n$.
Also, for any $i$ and $j$, by simple conditioning,
we have 
$$
\EXP \left\{ (N_i - \EXP N_i) (N_j - \EXP N_j ) \right\} = 0.
$$
By Chebyshev's inequality, and the fact that $\Var \{ N_i \} \le b^2$,
we have
$$
\PROB \{ R \ge r \}
\le { b^2 |I(\epsilon)| \over \left( \sum_{i \in I(\epsilon)}  \EXP N_i -n \right)_+^2},
$$
where $(y)_+= \max (u,0)$.
Let $E_i$ be the event 
$$
\left[ \theta \sqrt{n} \le Q(T_i) \le {1 \over \theta} \sqrt{n} , \epsilon n \le T_i \le (1-\epsilon) n \right].
$$.
We have for $i \in I(\epsilon)$,
$$
\eqalignno{
\EXP N_i 
&\le \EXP \{ N_i \IND{G(\epsilon,\theta)} \} + b \PROB \{ G^c(\epsilon,\theta) \} \cr
&\le \mu_b (1+o(1)) \PROB \{G(\epsilon,\theta) \} + b \delta \cr
&\le \mu_b (1+o(1)) + b \delta, \cr }
$$
where the $o(1)$ refers to the behavior as $n \to \infty$ for
fixed $b, \epsilon, \theta$.
Similarly,
$$
\eqalignno{
\EXP N_i 
&\ge \EXP \{ N_i \IND{E_i} \}  \cr
&\ge \mu_b (1+o(1)) \PROB \{ E_i \}  \cr
&\ge \mu_b (1+o(1)) ( 1 - \PROB \{ G^c(\epsilon,\theta)  \}  ) \cr
&\ge \mu_b (1+o(1)) ( 1 - \delta ). \cr }
$$
Therefore,
$$
\eqalignno{
\mu_b (1+o(1)) ( 1 - \delta ) (r - 2 \epsilon n)
&\le \mu_b (1+o(1)) ( 1 - \delta ) |I(\epsilon)|  \cr
&\le \sum_{i \in I(\epsilon)} \EXP N_i \cr
&\le \left( \mu_b (1+o(1)) + b \delta \right) |I(\epsilon)|. \cr
&\le \left( \mu_b (1+o(1)) + b \delta \right) r. \cr}
$$
For $n$ large enough,
$$
\sum_{i \in I(\epsilon)}  \EXP N_i -n \ge \mu_b ( 1 - 2 \delta ) (r - 2 \epsilon n)  -n.
$$
To show that for any small $\gamma>0$, $\PROB \{ R \ge n(1+\gamma)/\mu_b \} \to 0$,
we choose first $\epsilon = \gamma/(4\mu_b)$, then $\delta = \gamma / (4 + 4\gamma)$,
and then $\theta > 0$ so small that $\PROB \{ G^c (\epsilon, \theta) \} \le \delta$.
With those choices,
$$
\eqalignno{
\mu_b ( 1 - 2 \delta ) (\mu_b^{-1}(n + n\gamma) - 2 \epsilon n)  -n
& = n ( 1 - 2 \delta ) (1 + \gamma/2 )  -n \cr
& = n {\gamma^2 \over 4+4\gamma } \cr
& \isdef n \gamma'. \cr}
$$
So, for $n$ large enough, by Chebyshev's inequality invoked
above,
$$
\PROB \{ R \ge n(1+\gamma)/\mu_b \}
\le  {b^2 r \over   (n \gamma' )^2 } \to 0.
$$

The other side is dealt with in the same manner. First,
$$
\eqalignno{
\PROB \{ R < r \}
&\le \PROB \left\{ \sum_{i \in I(\epsilon)} N_i \ge n \right\} \cr
&= \PROB \left\{ \sum_{i \in I(\epsilon)} (N_i-\EXP N_i) \ge n - \sum_{i \in I(\epsilon)}  \EXP N_i \right\} \cr
&\le { b^2 r \over \left( n - \sum_{i \in I(\epsilon)} \EXP N_i \right)_+^2  }. \cr}
$$
Note that for $n$ large enough,
$$
n - \sum_{i \in I(\epsilon)}  \EXP N_i
\ge n  - \left( \mu_b + 2 b \delta \right) r.
$$
To show that for any small $\gamma>0$, $\PROB \{ R < n(1-\gamma)/\mu_b \} \to 0$,
we choose first $\delta = \gamma \mu_b / (2b)$,
and then $\theta > 0$ so small that $\PROB \{ G^c (\epsilon, \theta) \} \le \delta$.
With those choices,
$$
n  - \left( \mu_b + 2 b \delta \right) {n(1-\gamma) \over \mu_b}
= n \gamma^2.
$$
So, for $n$ large enough, by Chebyshev's inequality invoked
above,
$$
\PROB \{ R < n(1-\gamma)/\mu_b \}
\le  {b^2 r \over   n^2 \gamma^4 } \to 0.
$$

\end{plain}

\section*{6~Empirical results and discussion}

In the spirit of the remark by Hooker quoted in the introduction,
we put the result of Theorem 1 to the test experimentally. 
Except for $\Delta=2$ we considered critical Galton-Watson trees with
$p_i = 1/{i\Delta}, i=1,2,...,\Delta$ and $p_0= 1-\sum_{i=1}^{\Delta} p_i$.
For this family of trees we have that $\sigma^2=(\Delta - 1)/2$.
For $\Delta=2$ we used the distribution $p=(1/3,1/3,1/3)$ which has $\sigma^2=2/3$.
For each $\Delta =2,3,5,10,20,40$ we generated random Galton-Watson trees until we found one with at least
$10^8$ nodes. 
In Table \ref{tab:results} we show
the results of searching these trees with various budgets $b$. As before we 
denote the number of restarts by $R_n$. According to Theorem 1, $R_n/\sigma n$ should
asymptotically approach $\sqrt{\pi / 8b}$ as $b$ gets large. The results in Table \ref{tab:estimates} show
good empirical agreement with this asymptotic bound even for small values of $b$.

\begin{table}[h!tbp]
\centering
\scalebox{0.9}{
\begin{tabular}{|c | r r r r  r  r| }

\hline
Budget & \multicolumn{6}{c|}{$R_n$}  \\
$b$ &  $T_2~~~~$ & $T_3~~~~$ & $T_5~~~~$ & $T_{10}~~~~$  & $T_{20}~~~~$  & $T_{40}~~~~$ \\
\hline
50 &36164525 & 70861643 & 51881667& 52454166 & 118688002  & 51764031  \\
500& 10656708& 21596567 &16331233&17384235  & 40853820  & 18621338  \\
5000&3293174&6764700&5181648&5584327  & 13295349  &  6111917   \\
50000&1029919&2121621&1643866&1758032   & 4242028  & 1994306 \\
500000&321993&675660&520932&572507   &  1340837 &  620958 \\
\hline
$n$ & 440480727& 722813312& 393789282 & 264933315 & 434056761  & 145049024 \\
\hline
seed & 1486862923 & 2053907278 & 2018277783 & 1671700366 & 1992061106  & 2121539929 \\
\hline
\end{tabular}
}
\caption{Test results}
\label{tab:results}
\end{table}

\begin{table}[h!tbp]
\centering
\scalebox{0.9}{
\begin{tabular}{|c | c c c c  c  c  |c|}

\hline
Budget &  \multicolumn{6}{c|}{$R_n / \sigma n$}   & Estimate \\
$b$ &  $T_2$ & $T_3$ & $T_5$ & $T_{10}$ & $T_{20}$ & $T_{40}$ & $\sqrt{\pi / 8b }$ \\
\hline
50 & .10055 & .09804   & .09316 & .09333 & .08872  & .08082  & .08862     \\
500 &  .02963 & .02988 & .02933 & .03093 & .03054   & .02907  & .02802        \\
5000& .00916& .00936 & .00936 & .00994 & .00994  &  .00954 & .00886      \\
50000&  .00286 & .00294 & .00295 & .00313   & .00317  & .00311   & .00280    \\
500000& .00090& .00093 & .00094 & .00102 &  .00100 & .00097 & .00089     \\
\hline
\end{tabular}
}
\caption{Accuracy of Theorem 1 estimate}
\label{tab:estimates}
\end{table}

We now discuss how results like Theorem 1 are useful in practice. Recall
from the introduction that budgeting causes two competing forms of overhead:
restart cost and idle time. Let us model
a unit of computation time by a single node evaluation in Algorithm \ref{alg:bdfs}. We can
model the restart cost for a job as a constant $r$ units of time, i.e., $r$
node evaluations. For example a node evaluation in \mplrs involves pivoting
an $m$ by $n$ matrix. The restart cost in this case is approximately $r=n$,
as the input matrix has to be pivoted up to $n$ times to reach a restart cobasis.
Note that the restart cost is born by an individual worker and so the
total restart cost is independent of the number of workers.
As a proportion of the total work required it is precisely the ratio $R_n/n$ that
is estimated in Theorem 1.

If the job list becomes empty then any worker returning for work remains idle
until some new jobs arrive. The cost is the number of idle workers
for each time unit when the list is empty. Unlike restart costs, these costs
scale up as the number of workers increases and can greatly reduce the
parallelization efficiency achieved. To minimize these costs the job list should always
be well populated.

In practice the budget is not fixed for the duration
of a run. The master monitors the size of job list and if it drops too low scales
the budget down and if it gets too large scales the budget up. 
Theorem 1 tells us how this scaling will work for Galton-Watson trees: scaling down (up) by a factor
of 100 increases (decreases) the work returned on average by a factor of 10.
The dependence of the job list size on the square root of the budget explains
why relatively small changes in the budget do not greatly effect performance.
This can be observed in Figure \ref{fig:hist_sample} for example.

The empirical results obtained by Avis and Jordan (2015, 2016) come from
a variety of practical applications. Whilst these trees are not
generated according to the Galton-Watson model, the behaviour observed 
is surprisingly similar to that predicted by Theorem 1. 
It will be interesting to see if similar theoretical results hold for other 
classes of trees.

 \section*{7~Acknowledgements}
We gratefully acknowledge helpful conversations with 
Louigi Addario-Berry and Charles Jordan. The research of Avis was
supported by a JSPS
Kakenhi Grant and  a Grant-in-Aid for Scientific Research
on Innovative Areas, `Exploring the Limits of Computation (ELC)'.
The research of Devroye was supported by the Natural Sciences and Engineering Research Council of Canada.

 \section*{8~References}
\begin{plain}
\parskip=0pt 
\hyphenpenalty=-1000 \pretolerance=-1 \tolerance=1000 
\doublehyphendemerits=-100000 \finalhyphendemerits=-100000 
\frenchspacing 
\def\beginref{ 
\par\begingroup\nobreak\smallskip\parindent=0pt\kern1pt\nobreak 
\everypar{\strut}  } 
\def\endref{ 
\kern1pt\endgroup\smallbreak\noindent} 
\beginref

\endref

\beginref 
D.~Aldous,
``The continuum random tree. II. An overview,''
in: {\sl Stochastic Analysis (Durham, 1990)},
vol.~167,
pp.~23--70,
Cambridge Univ. Press, Cambridge,
1991.

\endref

\beginref 
D.~Aldous,
``The continuum random tree. I,''
{\sl The Annals of Probability},
vol.~19,
pp.~1--28,
1991.

\endref

\beginref 
D.~Aldous,
``Asymptotic fringe distributions for general families of random trees,''
{\sl The Annals of Applied Probability},
vol.~1,
pp.~228--266,
1991.

\endref

\beginref 
D.~Aldous,
``The continuum random tree. III,''
{\sl The Annals of Probability},
vol.~21,
pp.~248--289,
1993.

\endref

\beginref 
D.~Aldous
and J.~Pitman,
``Tree-valued Markov chains derived from Galton-Watson processes,''
{\sl Annals of the Institute Henri Poincar\'e},
vol.~34,
pp.~637--686,
1998.

\endref

\beginref 
K.~B.~Athreya
and P.~E.~Ney,
{\sl Branching Processes},
Springer Verlag,
Berlin,
1972.

\endref

\beginref
D.~Avis
and C.~Jordan,
``\mplrs: a scaleable parallel vertex/facet enumeration code,''
arXiv:1511.06487,
2015.

\endref

\beginref
D.~Avis
and C.~Jordan,
``A parallel framework for reverse search using \mts,'' 
arXiv:1610.07735,
2016.

\endref

\beginref 
J.~Bennies
and G.~Kersting,
``A random walk approach to Galton-Watson trees,''
{\sl Journal of Theoretical Probability},
vol.~13,
pp.~777--803,
2000.

\endref

\beginref
N.~Blumofe
and C.~Leiserson,
``Scheduling multithreaded computations by work stealing,''
{\sl Journal of the ACM},
vol.~46,
pp.~720--748,
1999.

\endref

\beginref 
T.~Duquesne,
``A limit theorem for the contour process of conditioned Galton-Watson trees. ,''
{\sl Annals of Probability},
vol.~31,
pp.~996--1027.,
2003.

\endref

\beginref 
M.~Dwass,
``The total progeny in a branching process,''
{\sl Journal of Applied Probability},
vol.~6,
pp.~682--686,
1969.

\endref

\beginref 
P.~Flajolet
and A.~Odlyzko,
``The average height of binary trees and other simple trees,''
{\sl Journal of Computer and System Sciences},
vol.~25,
pp.~171--213,
1982.

\endref



\beginref
J.~N.~Hooker,   
``Testing heuristics: We have it all wrong,''
{\sl Journal of Heuristics},
vol.~1,
pp.~33--42,
1995.

\endref


\beginref 
K.~Ito 
and H.~P.~McKean,
{\sl Diffusion Processes and their Sample Paths},
Springer-Verlag,
Berlin,
1974.

\endref

\beginref 
S.~Janson,
``Simply generated trees, conditioned Galton-Watson trees, random allocations and condensation,''
{\sl Probability Surveys},
vol.~9,
pp.~103--252,
2012.

\endref


\beginref 
D.~P.~Kennedy,
``The Galton-Watson process conditioned on the total progeny,''
{\sl Journal of Applied Probability},
vol.~12,
pp.~800--806,
1975.

\endref

\beginref 
D.~P.~Kennedy,
``The distribution of the maximum Brownian excursion,''
{\sl Journal of Applied Probability},
vol.~13,
pp.~371--376,
1976.

\endref

\beginref 
V.~F.~Kolchin,
``Branching Processes and random trees,''
in: {\sl Problems in Cybernetics, Combinatorial Analysis and Graph Theory (in Russian)},
pp.~85--97,
Nauka,
Moscow,
1980.

\endref

\beginref 
V.~F.~Kolchin,
{\sl Random Mappings},
Optimization Software Inc,
New York,
1986.

\endref

\beginref 
J.-F.~Le Gall,
``Marches al\'eatoires, mouvement Brownien et processus de branchement,''
in: {\sl S\'eminaire de Probabilit\'es XXIII},
edited by J.~Az\'ema, P.~A.~~Meyer and M.~~Yor,
vol.~1372,
pp.~258--274,
Lecture Notes in Mathematics,
Springer-Verlag,
Berlin,
1989.

\endref

\beginref 
J.-F.~Le Gall,
``Random trees and applications,''
{\sl Probability Surveys},
vol.~2,
pp.~245--311,
2005.

\endref

\beginref 
J.~F.~Marckert
and A.~Mokkadem,
``The depth first processes of Galton-Watson trees converge to the same Brownian excursion,''
{\sl Annals of Probability},
vol.~31,
pp.~1655--1678,
2003.

\endref

\beginref
T.~Mattson, B.~Saunders
and B. Massingill,
{\sl Patterns for Parallel Programming},
Addison-Wesley,
2004.

\endref

\beginref
C.~McCreesh
and P.~Prosser,
``The shape of the search tree for the maximum clique problem and the implications for parallel branch and bound,''
{\sl ACM Transactions on Parallel Processing},
vol.~2,
pp.~8:1--8:27,
2015.

\endref

\beginref 
A.~Meir
and J.~W.~Moon,
``On the altitude of nodes in random trees,''
{\sl Canadian Journal of Mathematics},
vol.~30,
pp.~997--1015,
1978.

\endref

\beginref 
J.~W.~Moon,
{\sl Counting Labelled Trees},
Canadian Mathematical Congress,
Montreal,
1970.

\endref

\beginref 
J.~Neveu
and J.~W.~Pitman,
``The branching process in a Brownian excursion,''
in: {\sl S\'eminaire de Probabilit\'es XXIII},
edited by J.~Az\'ema, P.~A.~~Meyer and M.~~Yor,
vol.~1372,
pp.~248--257,
Lecture Notes in Mathematics,
Springer-Verlag,
Berlin,
1989.

\endref

\beginref 
V.~V.~Petrov,
{\sl Sums of Independent Random Variables},
Springer-Verlag,
Berlin,
1975.

\endref

\beginref 
A.~R\'enyi 
and G.~Szekeres,
``On the height of trees,''
{\sl Journal of the Australian Mathematical Society},
vol.~7,
pp.~497--507,
1967.

\endref

\beginref
C.~Xu and
F,~Lau,
{\sl Load Balancing in Parallel Computers: Theory and Practice},
Kluwer Academic Publishers,
Norwell, MA,
1997.

\endref

\beginref 
\endref \bye 

\end{plain}

\end{document}